**Atomic-resolution study of oxygen vacancy ordering in La$_{0.5}$Sr$_{0.5}$CoO$_{3-\delta}$ thin films on SrTiO$_3$ during in-situ cooling experiments.**

Xue Rui and Robert Klie

Department of Physics, University of Illinois at Chicago, Chicago, IL, USA

The presence of oxygen vacancy, as well as ordering of vacancies plays an important role in determining the electronic, ionic and thermal transport properties of many transition metal oxide materials. Controlling the concentration of oxygen vacancies as well as the structures or domains of ordered oxygen vacancies has been the subject of many experimental and theoretical studies. In epitaxial thin films, the concentration of oxygen vacancies as well as the type of ordering depends on the structure of the support as well as the lattice mismatch between the thin films and the support. The role of temperature induced structural phase transitions on the oxygen vacancy ordering has remained largely unexplored. Here, we use aberration-corrected scanning transmission electron microscopy (STEM) combined with an in-situ cooling experiments to characterize the atomic/electronic structures of oxygen-deficient La$_{0.5}$Sr$_{0.5}$CoO$_{3-\delta}$ thin films grown on SrTiO$_3$ across the anti-ferrodistortive phase transition of SrTiO$_3$ at 105 K. We demonstrate that atomic-resolution imaging and electron energy-loss spectroscopy (EELS) can be used to examine variations in the local density of states as a function of sample temperature and thus of the structure of the support.

**I. Introduction**

Functional transition metal oxides have attracted a great deal of experimental and theoretical interest due to their wide range of properties, suitable for numerous practical applications, including fuel-cells, catalysts, thermoelectrics, spintronics, and nano-electronics. Perovskite oxides of the form ABO3 (B=Mn, Ti, Ni, Co) present a particularly interesting group of functional transition metal oxides, where the interplay of four degrees of freedom, lattice, orbital, charge, and spin generates a complex variety of ground states. An important attribute of this class of materials is that no single interaction or degree of freedom dominates. Consequently, subtle changes of external and internal parameters, such as temperature, electric or magnetic fields, strain at interfaces or defects, or ionic radii due to dopants may tremendously affect the overall properties of the system. Fascinating transport properties, such as high-temperature superconductivity, [1, 2] thermoelectric,[3, 4] ferro-electric,[5, 6] proton-,[7, 8] or magneto-transport[9, 10] are related to such induced phase transitions between different ground states and types of ordering.

It has become increasingly clear that oxygen vacancy ordering play an important role in controlling multiple physical property, such as ferroelectricity[11, 12], electronic[13], ionic[14] or thermal transport[15] as well as magnetic property.[16] In transition metal perovskite oxides, oxygen vacancies can formed ordered structures, such as the brownmillerite (BM) structure, where vacancies order in alternating planes along the *c*-axis forming fully stoichiometric oxygen octahedral and oxygen deficient tetrahedral layers. This ordering of vacancies does not only give rise to charge disproportionation of the transition metal ions on the *B* site, it also causes periodic changes in the perovskite lattice constant. This competition between the interlayer separation between tetrahedral planes and tetrahedral chain distortion within tetrahedral plane results in diversities of BM domains of different symmetries.[17] Furthermore, this ordered oxygen vacancy can also interact with the ferroelastic distortions, lowering the lattice stability and inducing the polar behavior within a given heterostructure. [18,19]

Oxygen vacancies ordering resulting in the formation of BM structures has been reported in oxygen-deficient $SrCoO_{3-\delta}$,[20] Sr-doped $LaCoO_{3-\delta}$ (LSCO) poly-crystalline bulk[21] and thin films[22,23], as well as other perovskite Co-oxides.[24] In $La_{1-x}Sr_xCoO_{3-\delta}$, where x>0.5, the formation of oxygen vacancies is enabled by the instability of $Co^{4+}$ in octahedral coordination, preferring a conversion to $Co^{2+}/Co^{3+}$ while releasing oxygen[25]. In thin film form, the lattice mismatch between epitaxial LSCO thin films and the substrate can result in biaxial strain stress, which will affect the domain orientation of oxygen vacancies in the BM structure. For compressively strained films, as found in LSCO films grown on $LaAlO_3$, the oxygen vacancy ordering occurs with an orientation that is parallel to the substrate-film interface; tensile strained thin films show an increased in-plane lattice parameter, which enabling oxygen vacancy ordering perpendicular to the interface.[26] Based on these observations, biaxial strain modulation can be used to control the oxygen vacancy ordering structure, and thus modifying the functionality of the thin film.[27, 28]

In addition to the interfacial strain modulation of thin film, coupling of interfacial oxygen octahedral tilts between the substrate and the thin films can also be used to modify the films' electronic/magnetic behavior. In perovskite epitaxial thin film, the physical properties depending on electronic bandwidth of transition metal *B* have been demonstrated to directly couple to octahedral rotations, since that the octahedral tilt angle is associated with B-O-B bonding angle and the length via overlap of the *d* orbitals and localized charge carriers.[29]

Aberration-corrected scanning transmission electron microscopy imaging techniques, such as annular bright-field (ABF) imaging, which can directly map the oxygen sublattice, has been utilized to directly map the $BO_6$ tilt pattern with pico-meter resolution,[30] as well as the presence or orientation of oxygen vacancy ordered BM domains.[20, 24, 31-35,36] Using accurate and quantitative mapping of B-O-B angle in ABF images in combination with DFT calculation and neutron diffraction measurements, Lee *et al.* have demonstrated that properties of multiferroic $BiFeO_3$ thin films can be artificially manipulated via changing oxygen octahedral tilt angles using strong oxygen octahedral coupling with $SrRuO_3$ layers. [37]

Recent studies have demonstrated that the magnetic properties inside LSCO thin film are affected by the choice of substrate, such as $NdGaO_3$ and $(LaAlO_3)_{0.3}(Sr_2AlTaO_6)_{0.7}$ (LSAT). While these substrates which have nearly same lattice spacing (3.861Å for $NdGaO_3$ and 3.868Å for LSAT), they exhibit different oxygen octahedral tilting pattern ($a^-a^-a^+$ for $NdGaO_3$ and $a^0a^0a^0$ for LSAT), which appears to influence the ferromagnetic ordering in epitaxial thin films.[22]

In this paper, we examine the effects of structural phase transitions of the thin film substrate on the oxygen vacancy ordering and electronic structure of epitaxial $La_{0.5}Sr_{0.5}CoO_{3-\delta}$ (LSCO) thin films. Specifically, $SrTiO_3$ (STO) is known to undergo an antiferrodistortive (AFD) phase transition from cubic structure to tetragonal structure at 105 K, which is due to the antiphase rotation of $TiO_6$ oxygen octahedral around one of the unit cell axis.[38] Using in-situ sample cooling combined with atomic-resolution imaging and EEL spectroscopy, we characterize the degree of coupling between the rotated $TiO_6$ and $CoO_6$ octahedra at the LSCO/STO interface. LSCO films grown on $LaAlO_3$ (LAO) substrates are used as a reference sample to distinguish the effects of interfacial strain and the STO structural phase transition at cryogenic temperature (95K).

The results reported in this paper are organized as follows: We first focus on atomic structure characterization of LSCO on STO and LAO at both room temperature (300 K) and cryogenic temperature (95 K). Then, we present atomic-resolution EELS analysis of the ordered oxygen vacancies for both samples at 300 K and 95 K.

## II. Experimental Details

$La_{0.5}Sr_{0.5}CoO_{3-\delta}$ epitaxial films with thickness of 12 nm were deposited on $SrTiO_3$ (001) substrate and $LaAlO_3$ (001) substrate respectively by high pressure reactive DC magnetron sputtering (more details can be found here[39, 40]). Atomic resolution scanning transmission electron microscopy (STEM) images were acquired using the probe aberration-corrected JEOL JEM-ARM 200CF at UIC equipped with cold field emission gun operated at a primary electron energy of 200kV. Electron energy loss spectrum (EELS) characterization was conducted using the Gatan Enfina and Gatan Quantum GIF spectrometers. The convergence semi-angle for STEM images and EELS was 13.4 mrad, the collection semi-angle for HAADF, LAADF and ABF detectors were set at range of 90-370 mrad, 40-160 mrad and 11-25 mrad, respectively. The collection semi-angle for EELS is 50 mrad. The corresponding probe current is 11 pA. The cooling experiments were conducted using Gatan 636 double-tilt liquid nitrogen (LN$_2$) cooling holder. TEM samples were prepared by traditional cross-section sample preparation method, using Multiprep wedge polishing followed by Ar$^+$ ion milling with the Fishione 1050 ion mill, and nano-milling by Fishione 1040 nano mill. For STEM imaging at 95 K, drift rates below <0.5 Å/s can be achieved, allowing for the acquisition of images with 1024 x1024 pixels. The HAADF images shown here were filtered using the HREM Filter plugin within Gatan's Microscopy Suite (GMS). ABF images are filtered with Principal Component Analysis. Atomap python package was used to determine the atom positions.[41] Low-dose EELS linescans or maps were acquired with 0.1s/pixel acquisition. All the final spectrums are aligned using the La M-edge at 832 eV and smoothed by Gaussian function within GMS.

## III. Results

First, the structure of oxygen vacancy ordering domains is characterized at LN$_2$ temperature (95K) and compared to structures found at room temperature using atomic-resolution HAADF imaging. Figure 1 shows atomic-resolution HAADF STEM images of La$_{0.5}$Sr$_{0.5}$CoO$_{3-\delta}$ on SrTiO$_3$ (STO) and LaAlO$_3$ (LAO) substrates in the [1 0 0] orientation at 300 K and 95 K, respectively. Since the contrast of HAADF images is directly proportional to the average atomic number of the imaged elements, direct imaging of oxygen atoms is not possible in this projection. The La/Sr in columns can be seen as the highest intensity contrast in Figure 1. The most noticeable feature of the LSCO film are dark and bright stripes, visible at both room temperature and LN$_2$ temperature, which has been previously reported as the brownmillerite structure (BM), where the oxygen vacancies order in alternating oxygen octahedral and tetrahedral planes along the (001) direction.[23, 35] This superlattice feature appears perpendicular to the interface in LSCO films on STO, while it is parallel for LSCO films on LAO substrate in agreement with previous reports.[26] Furthermore, it can be seen that the dark and bright columns are not directly touching the substrate/film interface, which is due to the tendency of oxygen vacancies toward clustering near the surface instead of interface. As Petrie *et al.*[42] and Tahini *et al.*[43] reported in their studies of in-plane biaxial strain effects on the formation of oxygen vacancies in $SrCoO_{3-\delta}$ thin film, the tensile strain stress at the surface lowers the oxygen vacancy formation energy and facilitates the oxygen vacancy ordering to form near the film surface. In addition, the epitaxial lattice spacing of LSCO results in a strained thin film forcing the oxygen vacancy to disorder at the LSCO/STO interface.[44] It appears that the critical lengths scale for such an effect is of the order of 4-5 unit cells.

As far as the lattice parameters for the BM structure is concerned, the lattice spacing of the bright and dark layers in the LSCO films are measured from images similar to those shown in Figure 1 and the average values are summarized in Table I. At room temperature, we find that the lattice spacing for bright layer is smaller than dark layer, as previously reported by Gazquez *et al.*[26] After cooling down to low temperature around 95K, the lattice spacing for bright layer is reduced while the dark layer is enlarged for LSCO film on STO. The lattice spacing remains unchanged in LSCO film on LAO, providing an initial suggestion that the STO phase transition affects the crystal structure of the LSCO film near the substrate/film interface.

In addition to the BM-type structure of oxygen vacancy ordering, many other forms of ordered vacancies have been suggested in LSCO.[45] Therefore, we use the ABF image, which is sensitive to light elements, such as oxygen, to directly image the oxygen coordinate environment in the LSCO thin films. The pseudo cubic [1 1 0] or [1 -1 0] projection provides the best viewing direction to establish the oxygen vacancy ordering type, since the oxygen atomic columns are clearly visible as shown in Figure 2 (a) and (b). For LSCO film on STO, we focus on an area where the oxygen vacancy ordering domain are parallel to the interface and images the oxygen vacancy ordering in the [1 1 0] direction. Figure 2(a) shows the LSCO film on STO, where the shift of the Co atomic columns as the result of missing oxygen columns is clearly visible. The proposed structure for the BM phase is also shown in this projection, which agrees with the structure observed in Figure 2(a). An ABF image of LSCO on LAO is shown as Figure 2(b), showing similar structural features and confirming that oxygen deficient LSCO films grown on either STO or LAO exhibit the BM-type oxygen vacancy ordering phase.

In the [1 0 0] projection, shown in the ABF image in Figure 2(c), the BM phase of the LSCO thin film appears to be distorted, potentially due to the strain imposed by the STO substrate. Compared with standard BM structure in the same orientation, the oxygen atoms in the tetragonal $CoO_4$ column and the adjacent octahedral $CoO_6$ columns are shifted as indicated in Figure 2c. The LSCO films on LAO show the expected BM structure, as seen in Figure 2(d).

Next, the electronic structure for the LSCO/STO heterostructure is examined as a function of temperature using EELS. The electron-energy-loss-near-edge-spectrum has been widely applied to characterize the *3d* orbital configuration and valence state in transition metal oxides. Here, EEL spectra of O *K*-edge and Ti *L*-edge are first acquired from the SrTiO$_3$ substrate at both room temperature and LN$_2$ temperature to probe the effects of the STO phase transition on the EELS near edge fine-structure (ELNES). As shown in Figure 3(a), both O K-edge spectra at 300 K and 95 K exhibit four prominent peaks, marked as *a, b, c, d*. The intensity of peak *a* has been associated with[46] hybridization of O *2p* and Ti *3d*, providing the information of Ti *3d* $t_{2g}$ and $e_g$ orbital configuration.[47] We find that there is no change O K-edge as a function of temperature. Figure 3(b) shows the Ti *L*-edge spectra at 300 K and 95 K. Here four peaks can be seen due to the splitting of Ti *3d* final states into the $t_{2g}$ and $e_g$ orbitals as well as the initial Ti $2p^{3/2}$ and $2p^{1/2}$ states.[48] Similar to the O $K$-edge ELNES, we do not find any change as a function of sample temperature. We proposed that core-loss EELS is not sensitive enough, or does not provide sufficient energy resolution to observe the changes in the electronic structure as the results of the STO phase transition at 105 K.

Figure 4 shows EEL spectra acquired from oxygen vacancy ordering domains in LSCO thin films grown on STO substrate and LAO substrate at both 300 K and $LN_2$ temperature. The Co $L_3$ and $L_2$ edges, which originate from the electron transition from $2p$ to unoccupied $3d$ states, can be used to determine the occupancy of the transition metal $3d$ orbitals. Particularly, Co $L_3/L_2$-ratio, known as white line ratio, has been widely utilized to determine the Co valence state.[49] As shown in Figure 4c), we do not find any significant change in the Co $L_3/L_2$ ratio at $LN_2$ and room temperature, indicating that the averaged Co valence remains unchanged as a function of temperature. This means that no additional vacancies are created as the result of cooling with prolonged electron beam exposure.

Figure 4(b) shows the O $K$-edge spectrum of LSCO films on STO substrate. Three prominent peaks represent different interaction between O and nearby orbital from Co and La/Sr. The O K-edge prepeak, labeled *a*, is associated with hybridization of O $2p$ and Co $3d$, peak *b* is due to interactions between the O $2p$ and La $5d$ orbitals and peak *c* stems from interaction between O 2p and La/Sr $4sp$.[50] The O K-edge prepeak *a* also includes peak $a_1$ (at 526 eV) and peak $a_2$ (at 527.5 eV), which is similar as the fine structure of the O K-edge pre-peak found in $SrFeO_{2.5}$ thin film.[44] Comparing the O $K$-edge spectra at 300 K and 95 K, the peak $a_2$ is prominent at 300 K but not at 95 K. Since Co valence state remains unchanged, this change in the O K-edge prepeak cannot be due to changes in the O stoichiometric. We propose that the change is due to a change in the Co $3d$ electronic orbital configuration. For reference, we perform similar experiments on LSCO thin films grown on LAO substrate, shown in Figure 4 (e) and (f). For the films grown on LAO, we do not find a similar temperature dependent change in the O K-edge prepeak. Hence, the change found in O K-edge prepeak in LSCO film on STO appears to be related to the STO substrate phase transition instead of other variables affected by temperature.

To determine the Co 3d orbital occupancy in the dark and bright layers of the LSCO thin films, atomic-column resolved EEL spectra were acquired at 300 K and 95 K. Figure 5(a) shows O *K*-edge for bright and dark layers in LSCO film on STO at 95K. The O K-edge pre-peak for the bright layers is located at 526 eV and at 527 eV in the dark layers. The Co *L*-edge spectra taken at 95K are shown in Figure 5(b). We determined that the Co white line ratio for bright layers is 3.0 while it is 3.8 in the dark layers. Using the relationship between Co white line ratio and Co valence state,[51] we find that the valence state for bright layer is Co 2.9+ and Co 2.5+ for the dark layers. This demonstrates that atomic column resolved EELS is possible at $LN_2$ temperature and can distinguish the Co valence state in the different layers of the BM-type oxygen-vacancy ordering. Column-resolved EELS result at 95K for LSCO thin film on LAO substrate is shown in Figures 5(c) and 5(d). Co L-edge spectra from LSCO films on LAO at 95K are shown in Figure 5(d). The measured Co white line ratio is 3.0 for bright layers while 3.9 in the dark layers, indicating a Co valence state of 2.9+ in the for bright layers and 2.4+ in the dark layer. Both LSCO film on STO and LAO exhibits the charge ordering in in the oxygen vacancy ordering area, which is similar to the Co valence state between dark and bright in $LaCoO_{3-\delta}$ thin film grown on $SrTiO_3$ substrate. [23]

Next, we compare EEL spectra acquired from LSCO films on STO at 300 K and 95 K, as shown in Figure 6. Figure 6(a) shows that O *K*-edge prepeak at 526 eV in the bright layers is suppressed at room temperature compared to $LN_2$ temperatures, while the intensity of prepeak at 527 eV in the dark layers is largely suppressed at room temperature. In Figure 6(b), Co L-edge appears to be mostly unchanged in both the bright and dark layers as a function of temperature, which demonstrates that the Co valence state/oxygen stoichiometry does not change significantly upon sample cooling. The reference data from LSCO films grown on LAO substrate is shown in Figure 7. Here, the O *K*-edge and Co *L*-edges remain unchanged in both the dark and bright layers at 300 K and 95 K. This comparison between the same thin film grown on different substrates further suggests that the variation found in the O K-edge prepeak of the LSCO/STO system is associated with changes in the STO substrate coupling into the LSCO layer rather than intrinsic changes in LSCO structure due to sample cooling.

**IV. Discussion**

Oxygen vacancy ordering in $La_{0.5}Sr_{0.5}CoO_{3-\delta}$ thin film on $SrTiO_3$ substrate has previously been studied by Gazquez *et al.*[52] who reported, using atomic-resolution STEM/EELS and DFT calculation, that in addition to the oxygen vacancies, the Co spin state also appears to orders in highly oxygen deficient areas of composition $La_{0.5}Sr_{0.5}CoO_{2.25}$ within $La_{0.5}Sr_{0.5}CoO_{3-\delta}$ thin film. According to the approximate linear relationship between lattice spacing and oxygen deficiency, $\delta$, proposed by Kim *et al.*,[22] we have determined $\delta$ in area where we performed our STEM/EELS analysis and found that the composition of our films ranges from $La_{0.5}Sr_{0.5}CoO_{2.7}$ to $La_{0.5}Sr_{0.5}CoO_{2.5}$, in good agreement with our average Co valence determination ranges from 2.9+ to 2.5+. We can, therefore, exclude Co spin state ordering at low temperature as the origin of the observed changes in the O K-edge fine-structure. Thus, we propose that changed in the O coordination in the octahedral and tetrahedral structure can explain the difference in the position of the oxygen K-edge prepeak between the bright and dark layers in the BM structure. That is also match with different energy level of Co 3d orbital in $CoO_4$ and $CoO_6$ as mentioned in work of Choi *et al.* [53].

To which extend the interfacial coupling influences the physical property of the heterostructure depends on the rigidity of thin film.[54] It has been demonstrated that $CoO_6$ exhibit weaker Jahn-Teller distortions compared to $MnO_6$ in $La_xSr_{1-x}MnO_3$ thin film by Sundaram *et al.* [55], which enables $CoO_6$ to be easier tilted by $TiO_6$ octahedral rotation, and to maintain the tilted pattern up to 10 nm from the interfaces, compared to decaying within several unit cells in $La_xSr_{1-x}MnO_3$. However, the tensile strain imposed by the STO substrate forces the LSCO lattice spacing to remain constant on average, the observed increase in lattice spacing of the oxygen deficient dark layers of the BM structure leads to a compensating decrease in interatomic distance of the bright layer. Furthermore, the twinning domains that form as the result of the phase rotation of the $TiO_6$ octahedra during the STO phase transition, as well as the different ligand fields between the $CoO_6$ octahedra and the $CoO_4$ tetrahedra make it possible for the bright and dark layers to respond differently at low temperatures.[17, 56] Finally, we have previously reported that the O *K*-edge prepeak is sensitive to Co $3d^6$ spin state transition in $LaCoO_3$.[57] However, in the LSCO films reported here, we do not expect a similar Co spins state transition and the mixed Co +2/+3 valence state make the determination of the Co ion spin state more difficult compare to $LaCoO_3$. Thus, additional DFT modeling will be required to distinguish the effects of valence and spin state transition in LSCO as a function of temperature.

**V. Conclusion**

We have demonstrated that atomic-resolution STEM-EELS analysis can be performed at both room temperature and $LN_2$ temperature. We have used this approach to directly measure the effects of the STO low temperature anti-ferrodistortive phase transition on the atomic/electronic structures of oxygen vacancy ordering in LSCO thin films. We find that both the lattices spacing and the O *K*-edge fine structures in dark and bright layers of the BM-type oxygen vacancy ordered domains respond in different ways during the cooling experiment only in films grown on STO. The same films grown on LAO do not show any temperature dependent changes in the atomic and electronic structures. Similar experimental approaches to atomic-resolution STEM imaging and EEL spectroscopy can now be used to explore structural or electronic transition in a large range of function transition metal oxide materials as function of temperature.

## VI. Acknowledgement

The authors want to thank Jeff Walter and Chris Leighton (University of Minnesota) for growing LSCO thin film samples. This work was supported by a grant from the National Science Foundation (Grant No. DMR-1408427). The acquisition of the UIC JEOL JEM-ARM200CF was supported by NSF MRI-R2 Grant (No. DMR-0959470). Support from the UIC Research Resources Center (RRC), in particular A.W. Nicholls and F. Shi is acknowledged.

**Fig 1.**

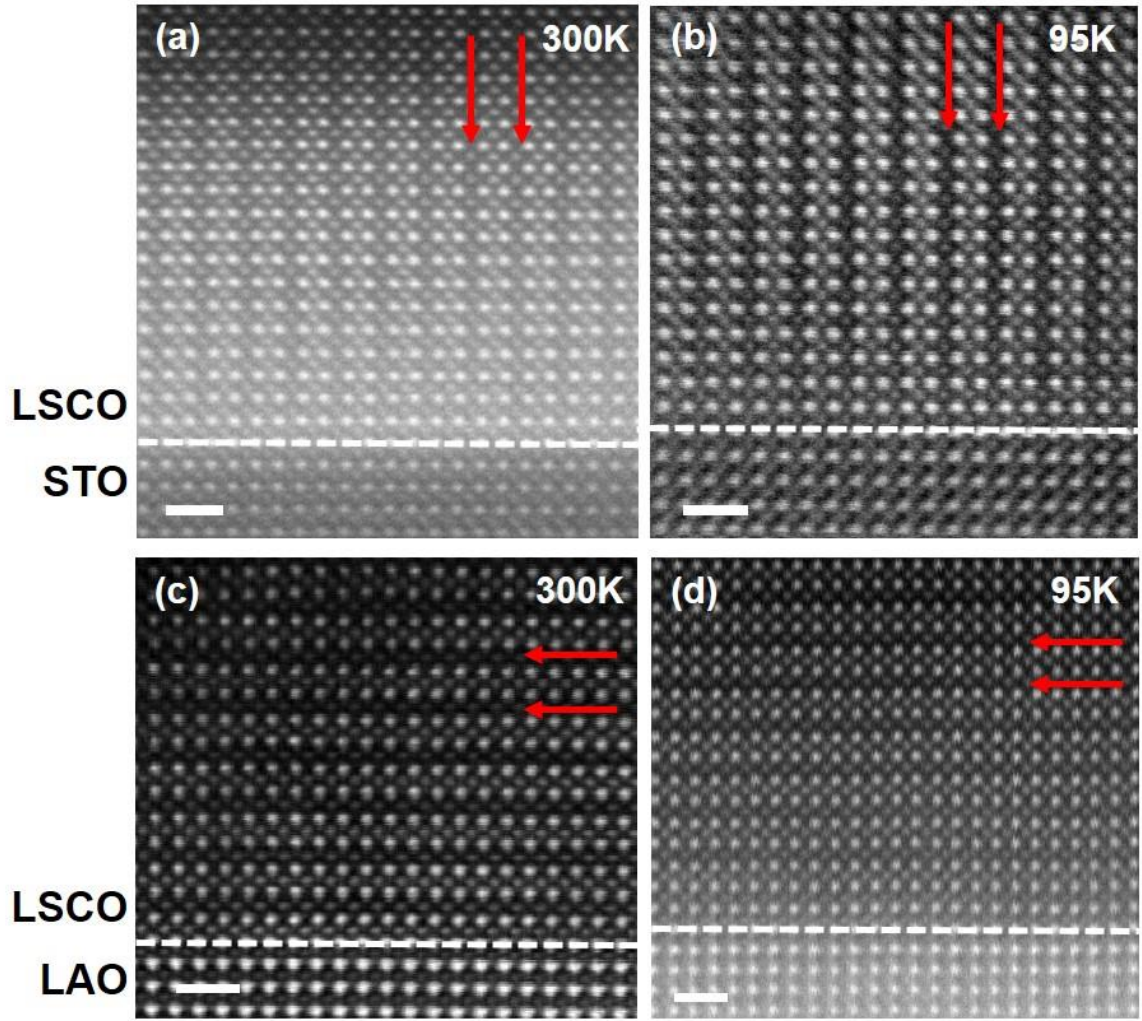

Fig. 1. (a) (b) Representative atomic-resolution HAADF image of LSCO thin film grown on STO substrate at 300 K and 95 K. (c) (d) Representative atomic-resolution HAADF image of LSCO thin film grown on LAO substrate at 300 K and 95 K, respectively. White arrows marked the dark/oxygen deficient layers. Scale bar is 1 nm.

**Table I**

| Average Lattice Spacing | | | |
|---|---|---|---|
| | | 300K (Å) | 95K (Å) |
| LSCO/STO $a_{STO}$=3.905 | Bright | 3.61±0.09 | 3.51±0.09 |
| | Dark | 4.29±0.11 | 4.33±0.11 |
| LSCO/LAO $a_{LAO}$=3.789 | Bright | 3.68±0.09 | 3.68±0.09 |
| | Dark | 4.33±0.11 | 4.32±0.10 |

Table I Statistical averaged lattice spacing measurement based on 10-20 HAADF images for each part.

**Fig.2**

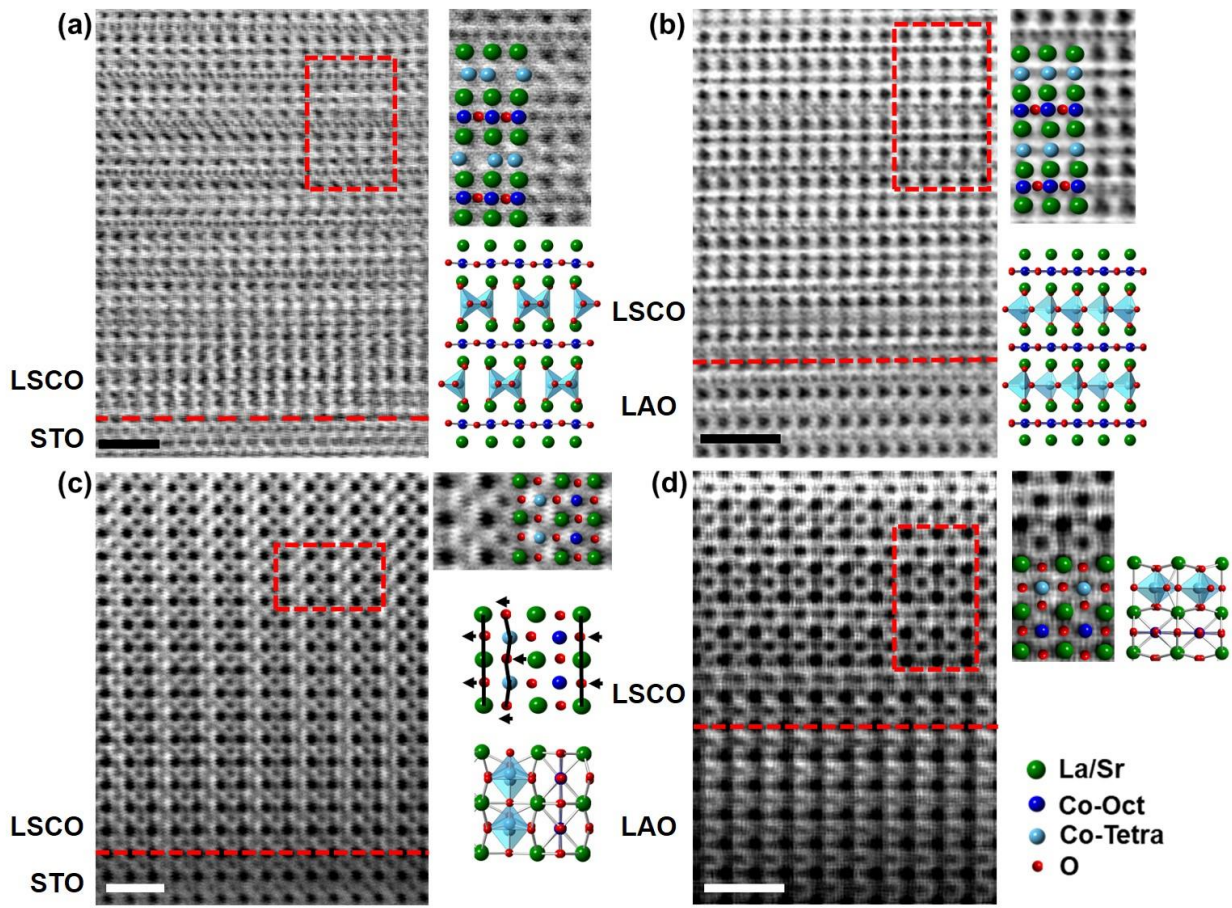

Fig. 2. (a) ABF image of LSCO thin film grown on STO substrate along [1 1 0]. The marked area shows the same structure as standard BM structure in [0 1 0] direction (b) ABF image of LSCO thin film grown on LAO substrate along [1 -1 0], the marked area shows the same structure as standard BM structure in [1 0 0] direction. (c) ABF image of LSCO thin film grown on STO substrate along [1 0 0]. The arrows mark the shift of oxygen atom compared with standard BM-structure (d) ABF image of LSCO thin film grown on LAO substrate along [1 0 0], the marked area shows the same structure as standard BM-structure. Scale bar is 1nm.

**Fig. 3**

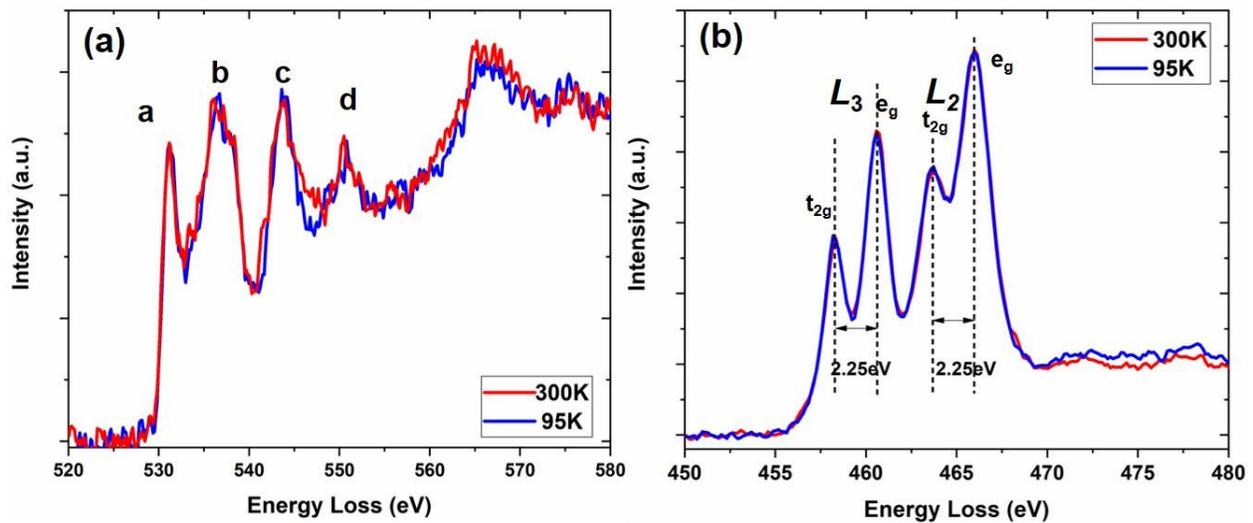

Fig. 3. (a) O *K*-edge acquired on SrTiO$_3$ substrate at room temperature and LN$_2$ temperature. (b) Ti *L*-edge acquired on SrTiO$_3$ substrate at room temperature and LN$_2$ temperature.

**Fig.4**

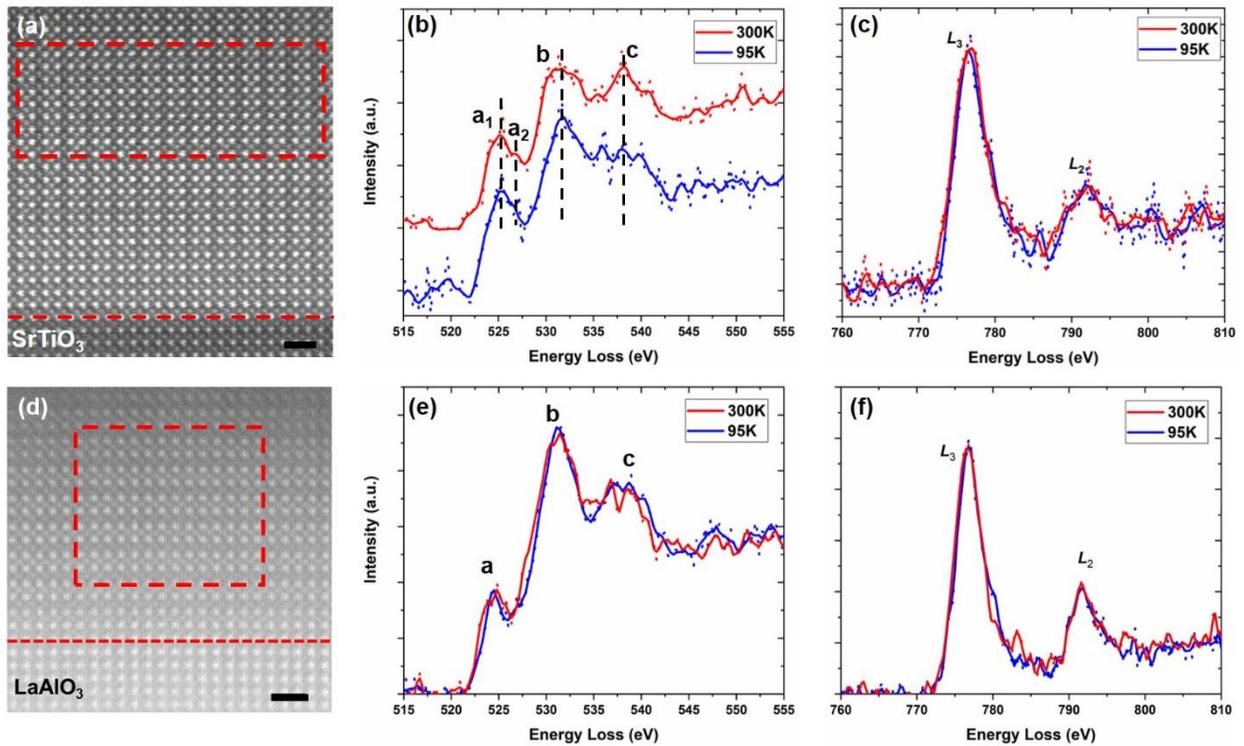

Fig. 4. (a) Representative HAADF image of LSCO on STO substrate acquired at room temperature. (b) (c) O *K*-edge and Co *L*-edge spectrum at 300K and 95K acquired at oxygen vacancy ordering domain area as marked in (a). (d) Representative HAADF image of LSCO on LAO at room temperature. (e) (f) O K-edge and Co L-edge spectrum at 300 K and 95 K acquired at oxygen vacancy ordering domain area as marked in (d). Scale bar is 1nm.

**Fig. 5**

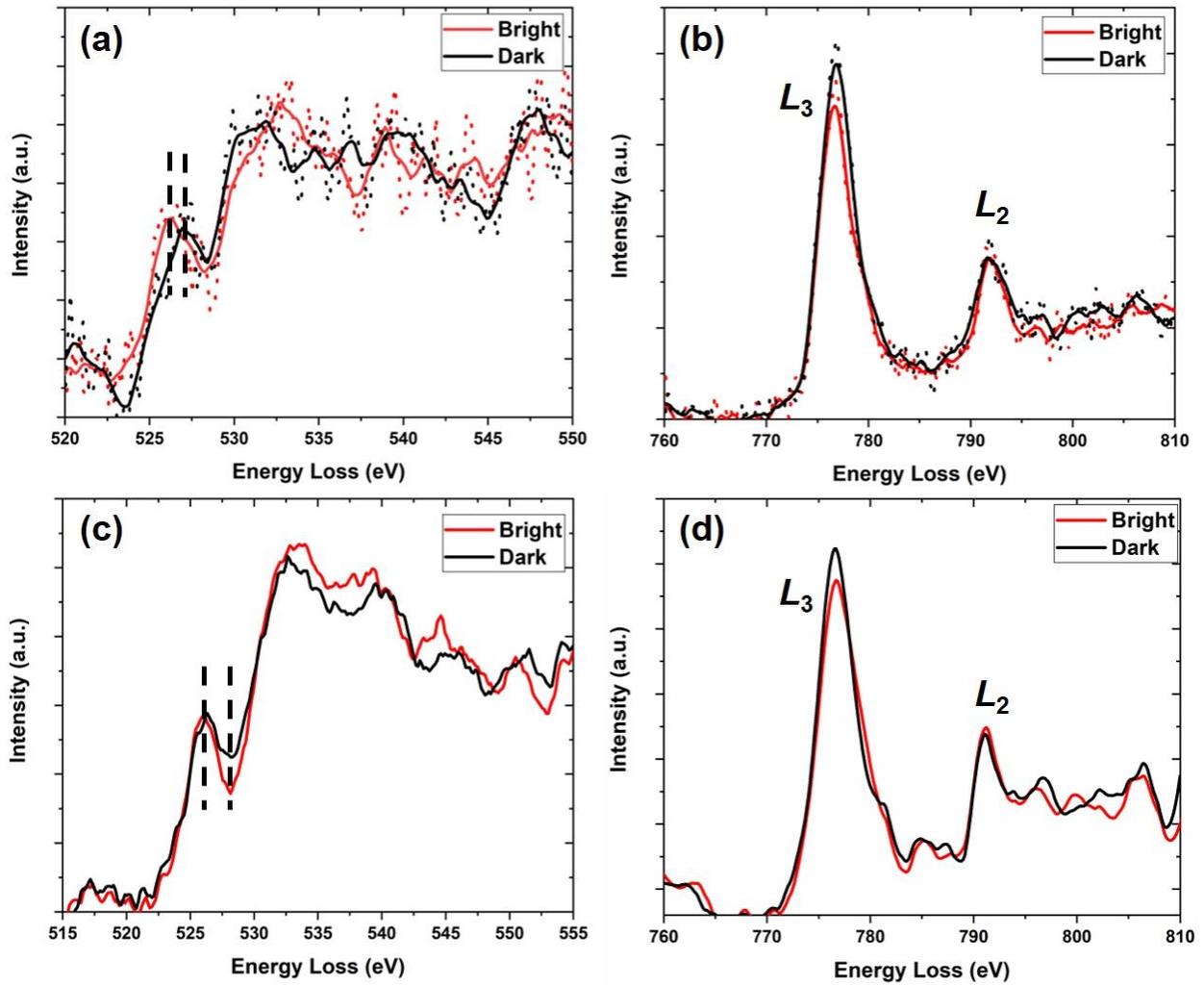

Fig. 5. (a) (b) O *K*-edge and Co *L*-edge spectrum extracted from dark and bright layer on LSCO grown on STO substrate at LN$_2$ temperature. (c) (d) O *K*-edge and Co *L*-edge extracted from dark and bright layer on LSCO grown on LAO substrate at 95 K.

**Fig. 6**

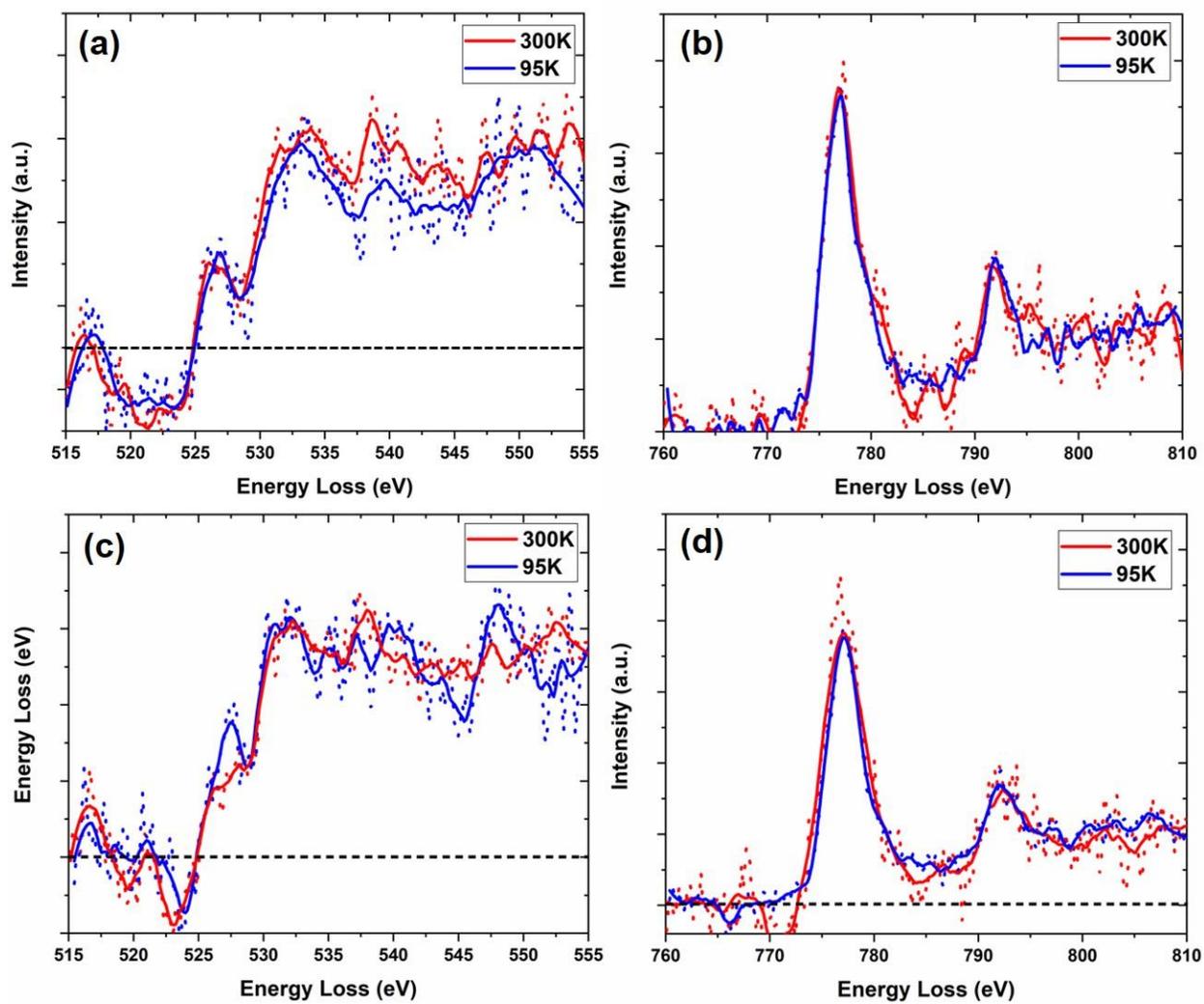

Fig. 6. (a-d) O *K*-edge and Co *L*-edge acquired at 300 K and 95 K from dark and bright layer respectively in LSCO on STO.

**Fig. 7**

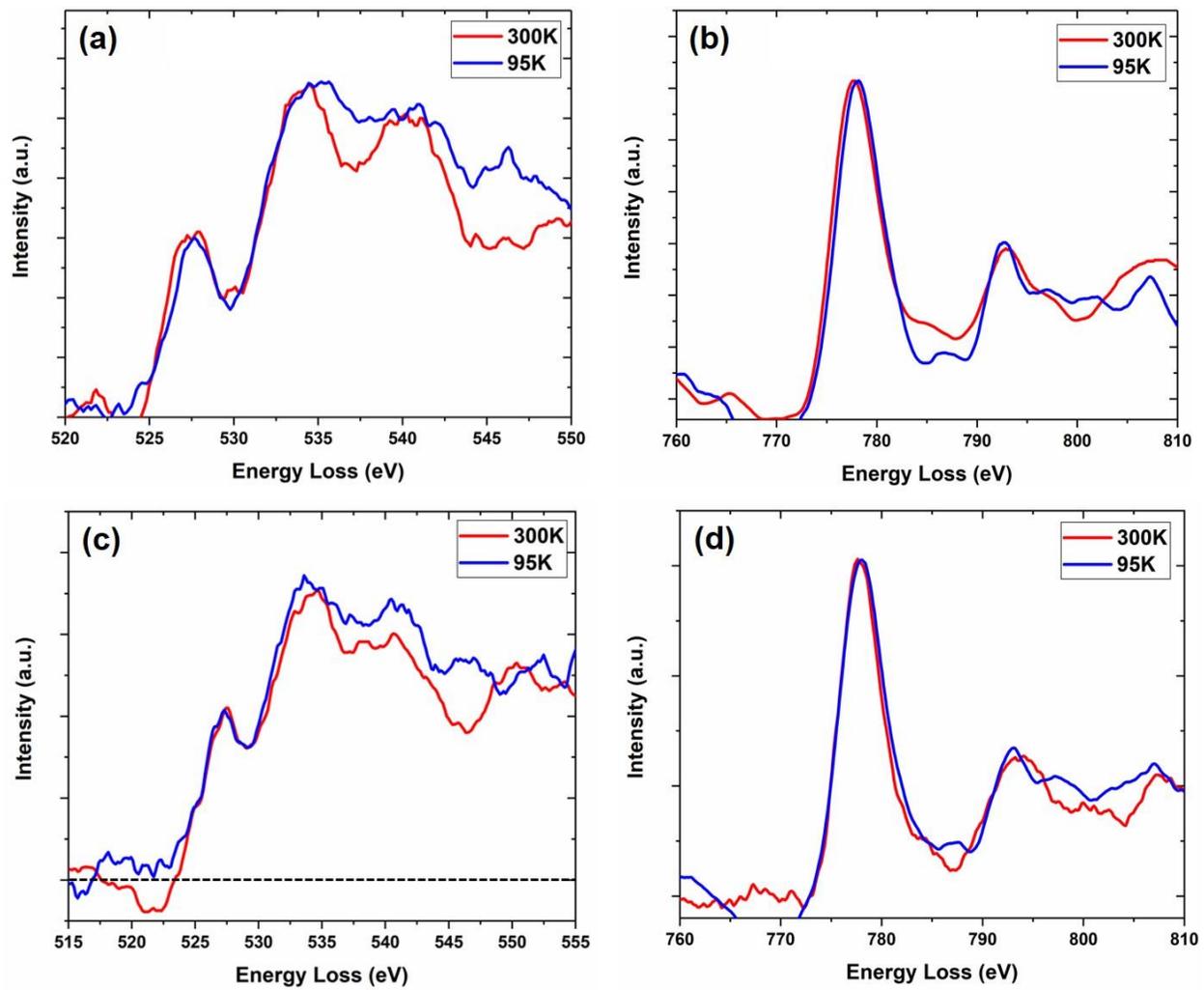

Fig. 7. (a-d) O *K*-edge and Co *L*-edge acquired at 300 K and 95 K from dark and bright layer respectively in LSCO on LAO.